\documentclass{nature}
\usepackage{color}
\usepackage{arydshln}
\usepackage{graphicx}  
\usepackage{lineno}
\usepackage{hyperref}
\hypersetup{
    colorlinks=true,
    linkcolor=blue,
    filecolor=magenta,      
    urlcolor=cyan,
}
\usepackage[final]{pdfpages}
\makeatletter
\let\saved@includegraphics\includegraphics
\AtBeginDocument{\let\includegraphics\saved@includegraphics}
\renewenvironment*{figure}{\@float{figure}}{\end@float}
\makeatother

\title{Primordial Earth mantle heterogeneity caused by the Moon-forming giant impact}

\author{Hongping Deng$^1$, Maxim D. Ballmer$^2$, Christian Reinhardt$^1$, Matthias M. M. Meier$^3$, Lucio Mayer$^1$, Joachim Stadel$^1$, Federico Benitez$^1$}

\begin{document}

\maketitle

\begin{affiliations}
 \item Center for theoretical Astrophysics and Cosmology, Institute for Computational Science, University of Zurich, Winterthurerstrasse 190, 8057 Zurich, Switzerland
 \item Institute of Geophysics, ETH Zurich, Sonneggstrasse 5, 8092 Zurich, Switzerland
 \item Institute of Geochemistry and Petrology, ETH Zurich, Clausiusstrasse 25, 8092 Zurich, Switzerland
\end{affiliations}

\begin{abstract}
The giant impact hypothesis for Moon formation\cite{Cameron1976,Canup2001} successfully explains the dynamic properties of the Earth-Moon system but remains challenged by the similarity of isotopic fingerprints of the terrestrial and lunar mantles\cite{Asphaug2014}. Moreover, recent geochemical evidence suggests that the Earth's mantle preserves ancient (or ``primordial'') heterogeneity\cite{Rizo2016,Mundl2017} that predates the Moon-forming giant impact\cite{Barboni2017}. Using a new hydrodynamical method\cite{Deng2017}, we here show that Moon-forming giant impacts lead to a stratified starting condition for the evolution of the terrestrial mantle. The upper layer of the Earth is compositionally similar to the disk, out of which the Moon evolves, whereas the lower layer preserves proto-Earth characteristics. As long as this predicted compositional stratification can at least partially be preserved over the subsequent billions of years of Earth mantle convection, the compositional similarity between the Moon  and the accessible Earth's mantle  is a natural outcome of realistic and high-probability
 Moon-forming impact scenarios\cite{Mastrobuono2017}. The preservation of primordial heterogeneity in the modern Earth not only reconciles geochemical constraints\cite{allegre1996,Mukhopadhyay2012,Rizo2016,Mundl2017} but is also consistent with recent geophysical observations\cite{Fukao2013,French2015,Jenkins2017,Waszek2018}. Furthermore, for significant preservation of a proto-Earth reservoir, the bulk composition of the Earth-Moon system may be systematically shifted towards chondritic values. 
\end{abstract}

 As the only planet in our solar system, the Earth is orbited by a single and massive moon. The leading theory for the formation of the Earth-Moon system with its high angular momentum involves a giant impact followed by lunar aggregation from the impact debris disk\cite{Cameron1976}. The canonical giant impact model involves a graze-and-merge impact, in which a Mars-sized body (or ``Theia'') collides with the proto-Earth at an oblique angle at roughly the escape velocity of the system\cite{Canup2001}. In this model, however, Theia contributes a larger fraction of silicates ($\sim$70\% by mass) to the proto-lunar disk than to the proto-Earth. Unless Theia and the proto-Earth had almost the same isotopic composition, this imbalance is at odds with the strong isotopic similarity of the Earth's and lunar mantles, e.g., in terms of oxygen\cite{Herwartz2014} and titanium\cite{Zhang2012}.

One way to reconcile this compositional similarity involves the post-impact re-equilibration of the Earth and the Moon-forming disk\cite{PahlevanStevenson2007}. This model, however, is unable to explain the isotopic similarity of the Earth and Moon in highly refractory elements, for example, titanium\cite{Zhang2012}. More recently, several alternative giant-impact models have been proposed. A near equal-mass ``Sub-Earth'' impact\cite{Canup2012} or the disruption of a fast-spinning Earth (close to self-breakup) by a small impactor\cite{Cuk2012} can indeed explain the isotopic similarity. However, the proposed solutions are low-probability events: equal-mass collisions are highly unlikely in the late stage of planetary accretion\cite{Mastrobuono2017}, and planetary embryos hardly reach spins that are close to self-breakup in hydrodynamic simulations\cite{Rufu2017}. Both models also predict an angular momentum that is too high for the early Earth-Moon system, and the mechanisms proposed to remove the excess angular momentum exclusively work in a narrow tidal parameter range\cite{Wisdom2015}. Alternatively, multiple-impact models have been suggested for lunar origin\cite{Rufu2017}, certainly broadening the favourable parameter space compared to single-impact models, and giving rise to mixing through mergers of moonlets with different isotopic composition. However, the dynamics of moonlets are highly uncertain, and primitive moonlets might be lost during repeated impacts pre-dating the Moon-forming stage \cite{Pahlevan2015}.

Here, we explore the mixing state of the Earth-Moon system in low-velocity, low-angular-momentum impact scenarios. We consider the canonical model\cite{Canup2001} and the hit-and-run model\cite{Reufer2012}(see Figure 1a, for these $v_{imp}<1.3v_{esc}$, where $v_{imp}$ is the
impact velocity and $v_{esc}$ is the escape speed), which are both high-probablity impact configurations \cite{Jackson2017} and lead result naturally in the current angular momentum of the
Earth-Moon system. We apply a new hydrodynamical Lagrangian method: Meshless Finite Mass (MFM) (see Methods subsection 2). This method is better at capturing fluid mixing and at resolving the core-mantle boundary than the widely-used smoothed-particle hydrodynamical approach\cite{Deng2017}.

\begin{figure}
  \includegraphics[width=\textwidth]{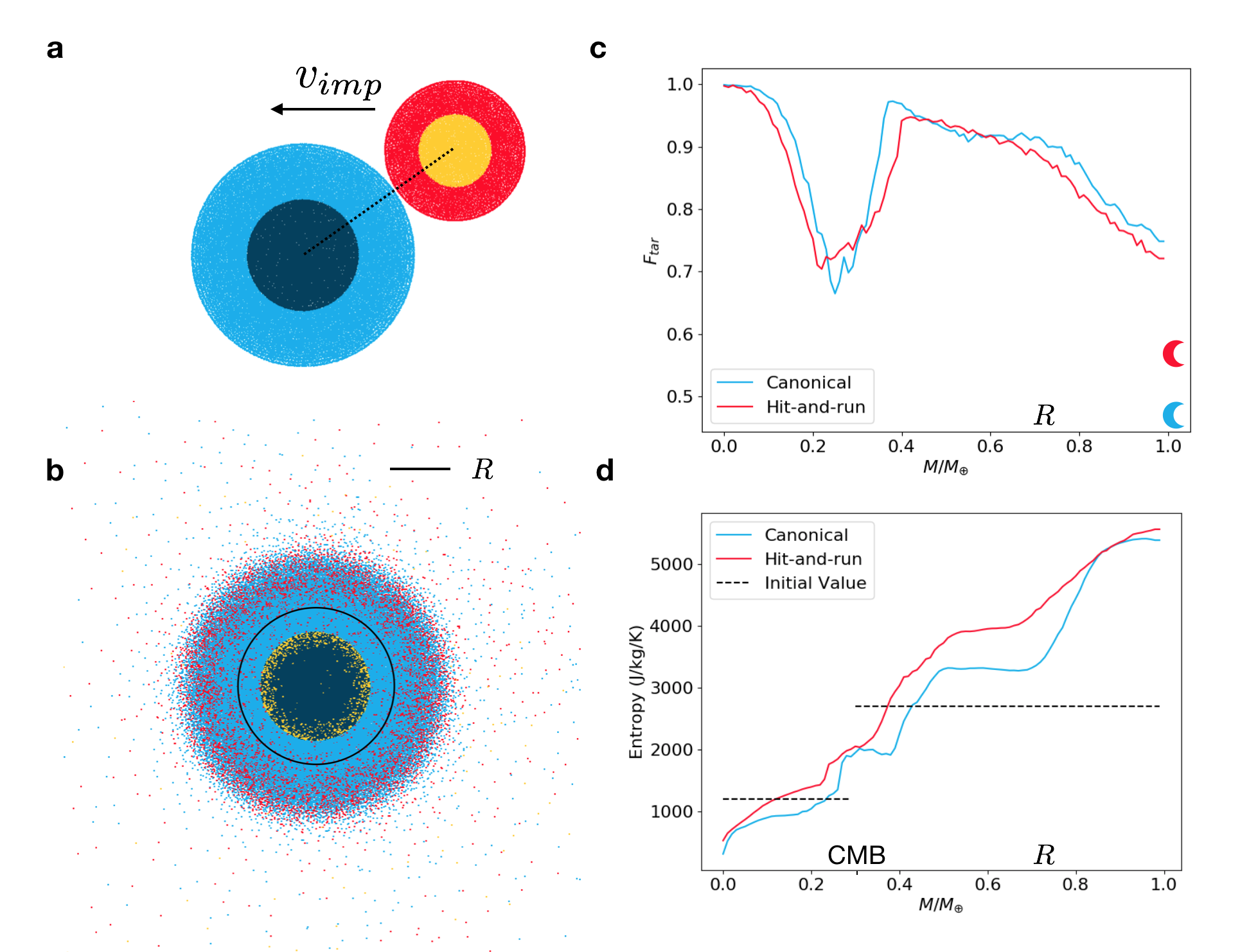}
  \caption{The internal structure of the post-impact Earth, after (i) canonical and (ii) hit-and-run collisions. (\textbf{a}) The initial condition just before the impact with the impactor's and target's mantle/core as marked with red/yellow and light-blue/dark-blue colors, respectively. The compositional structure of the post-impact target is visualized as (\textbf{b}) a 2D cross-section through the 3D model for run 13, and (\textbf{c}) 1D average profiles for runs 5 (canonical, blue) and 13 (hit-and-run, red). The average composition of the disk is denoted by a half-moon symbol. Panel \textbf{d} shows the related entropy profiles with black dash lines indicating the initial condition. A significant entropy jump is predicted at a radius $R$ (i.e., at a normalized enclosed mass of $\sim$0.7$M_{\oplus}$), corresponding to a kink in the compositional profile.\label{fig:1}}
\end{figure}

  In our simulations of such impacts, a strong shock propagates almost perpendicular to the line  connecting the centers of the two impacting bodies after the first contact. In both the canonical and hit-and-run scenarios, the part of the impactor that can avoid direct collision is sheared into a spiral structure which, shortly afterwards, collapses into clumps. The clumps re-impact the highly distorted target to eject some additional target material into the circum-planetary debris disk (Supplementary Video; Figure \ref{fig:1}b). Figure \ref{fig:1}c shows the predicted fraction of target material in the post-impact body, $F_{tar}$ as a function of normalized enclosed mass (which corresponds to the radius, within which a given planetary mass fraction is enclosed). Even though mixing is more efficient with our new method than in previous studies, most of the impactor's silicates still remain in the outer layer of the post-impact target mantle (low $F_{tar}$). This prediction is explained by inefficient transfer of angular momentum during the impact (Methods subsection 1). Focused shock heating in the outer layer (Supplementary Video) results in a steep entropy profile (Figure 1d) through most of the mantle. In particular, there is a distinct entropy jump at radius $R$, or at a normalized enclosed mass of $\sim$0.7 $M_{\oplus}$. In terms of mass, this $R$ corresponds to $\sim$1000 km depth in the present-day mantle.

For the metal core, our models also predict a steep entropy and compositional profile in the aftermath of the impact. The entropy of the deep core even drops below its initial condition of 1200 J/kg/K. This is well explained by the effects of phase transitions near the core-mantle boundary, which can be captured by the MFM method, and results in a redistribution of energy, and entropy, from the core to the mantle \cite{Deng2017}(Methods subsection 2, Extended Data Fig. 2, 6). The impactor's metal mostly remains near the top of the core. Depending on the conditions of metal-silicate (impactor metallic core/target silicate mantle) equilibration, this prediction may provide an explanation for the $\sim$300-km-thick compositionally stratified layer that is seismically observed at the top of the present-day outer core\cite{Helffrich2010}.

\begin{figure}
  \includegraphics[width=\textwidth]{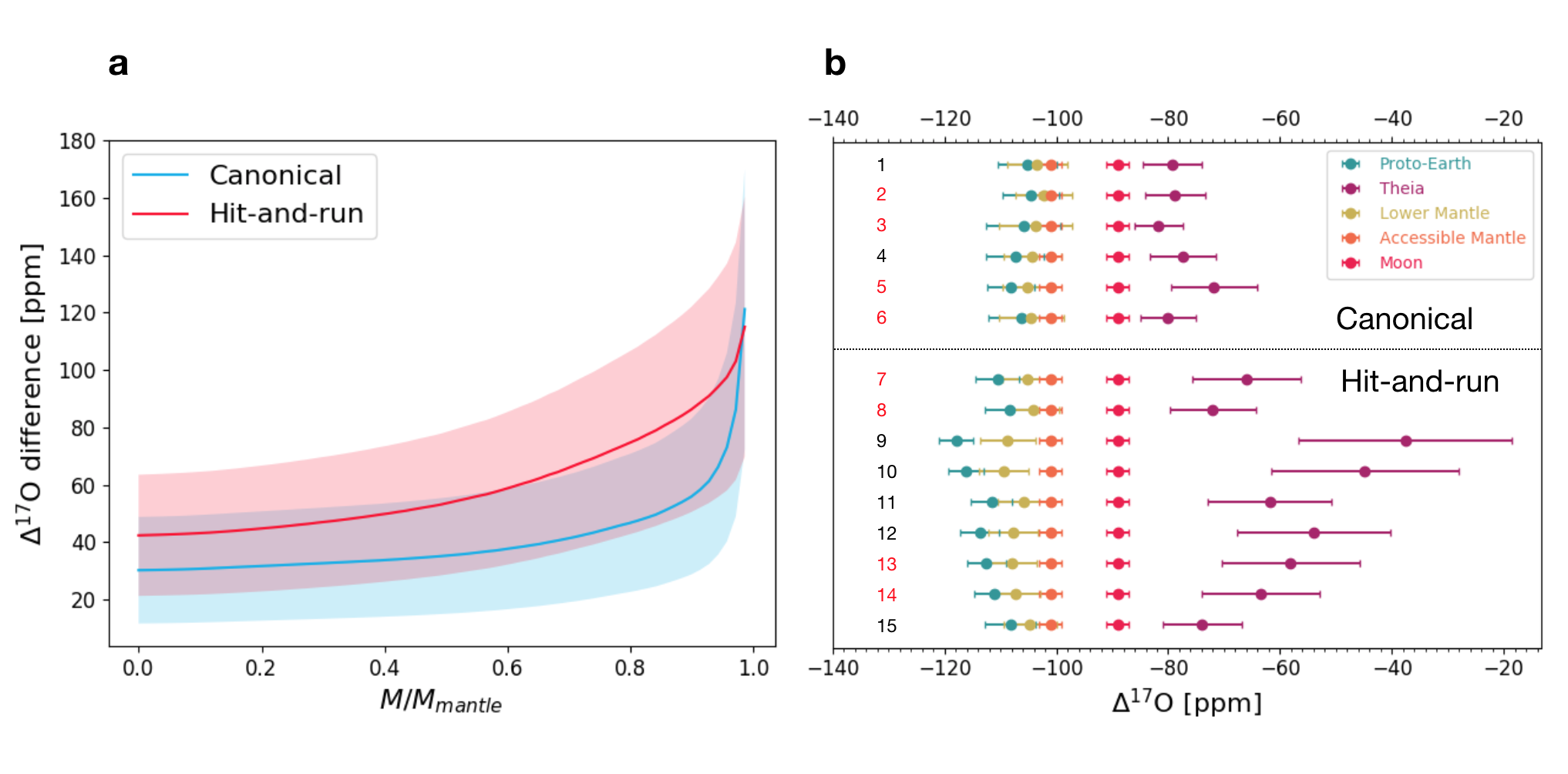}
  \caption{The oxygen isotopic composition of all reservoirs involved in the collision. Using the compositional profiles predicted by our models, we calculate the compositions of the lower-mantle layer, Theia and proto-Earth based on those of the accessible present-day mantle and Moon\cite{Herwartz2014}. Panel \textbf{a} shows the allowed $\Delta^{17}$O difference between Theia and the proto-Earth for runs 5 (blue) and 13 (red) as a function of the mass of the reservoir that remains unmixed with the accessible mantle. The shaded regions/error bars correspond to $1\sigma$ SEM uncertainty\cite{Herwartz2014}. Panel \textbf{b} shows the estimated $\Delta^{17}$O for all reservoirs and all runs (table \ref{tab:simulations}) assuming that no mixing occurs in the Earth's mantle across $R$ (i.e., mass of the unmixed reservoir is $\sim$50\% of that of the mantle). For example, Theia's oxygen isotopic composition could have been up to 54 ppm higher than that of the proto-Earth for run 13, simultaneously resulting in an isotopic difference between the accessible and lower mantle of 7 ppm. See Supplementary Material for details, as well as plots for elements Ti and Cr.\label{fig:oxygen}}
\end{figure}

To test our model predictions with geochemical observations, we estimate the unknown compositions of the impactor's (Theia) and target's (proto Earth) mantles from the known isotopic compositions of the accessible part of the Earth's mantle and Moon\cite{Herwartz2014} (See Supplementary material for calculation using alternative data). Figure \ref{fig:oxygen}a shows the allowed $\Delta {}^{17}$O-difference as a function of the mass of the mantle that remains poorly homogenized (or preserved) over the age of the Earth. If the present-day mantle is fully homogenized such that any primordial stratification is completely removed, a common assumption in previous studies, Theia and the proto-Earth must have been rather similar in composition. For example, only 30 ppm $\Delta^{17}$O difference between parent bodies are allowed in our best canonical model (run 5). On the other hand, if the assumption of full homogenization is relaxed, and the predicted mantle stratification can be (partially) preserved through the present day, larger compositional differences can be reconciled with the available data. For example, considering the preservation of a compositionally-distinct domain below $R$, differences of up to 54 ppm (run 13) in $\Delta {}^{17}$O are allowed (Figure \ref{fig:oxygen}b), particularly for the hit-and-run models, which display larger fractions of proto-Earth silicate material in the disk than the canonical models (table \ref{tab:simulations}; Figure \ref{fig:1}). These large values are consistent with the compositional difference between potential parent bodies of the Earth-Moon system in N-body simulations of planetary accretion\cite{Mastrobuono2017}. For realistic parent-body compositional differences\cite{Dauphas2017}, inefficient mixing of the Earth's mantle through time can indeed help to resolve the geochemical similarity of the accessible Earth mantle and Moon. Even for moderate mixing across $R$, our models can critically increase the allowed compositional difference between parent bodies (see Figure \ref{fig:oxygen}a).


Whether post-impact heterogeneity can persist through $\sim$1 Myr of magma-ocean and $\sim$4.5 Gyrs of mantle convection is controlled by the initial compositional and thermal profiles of the mantle. Our models predict the formation of a deep magma ocean due to the energy release of the giant impact\cite{Nakajima2015} (see Extended Data Fig. 6). While major-element compositions of the post-impact mantle layers above and below $R$ depend on the unknown bulk compositions of Theia and the proto-Earth, respectively, an enrichment of the lower (proto-Earth) layer in FeO and SiO$_2$ is generally consistent with the evolving of physical conditions of multi-stage core formation during progressive planetary accretion\cite{Kaminski2013,Rubie2015}. Furthermore, FeO-enrichment of the deep proto-Earth’s mantle may have been promoted by compositional fractionation (and subsequent overturn)\cite{elkins2008linked} during any magma-ocean episode(s) that predate(s) the Moon-forming impact. Even just a slight FeO-enrichment of the lower layer is sufficient to promote stable stratification through various magma-ocean stages. A long lasting stratification should be favoured by the entropy gradient across the post-impact mantle predicted by our models as the observed
sub-adiabatic entropy gradient should prevent redistribution by convection.

After the final magma-ocean episode that follows the giant impact, mixing during long-term solid-state mantle convection is controlled by the density and viscosity contrasts between the two layers. Primordial heterogeneity can survive mantle stirring as blobs over a range of spatial scales (meters to terameters)\cite{manga1996mixing,Ballmer2017}. Intrinsically high densities and viscosities of the primordial deep-mantle layer as sustained by an enrichment in FeO and SiO$_2$ of the proto-Earth's mantle (see above) impede efficient across $R$. For example, ref. \cite{Kaminski2013} predicts a molar Mg/Si of $\sim$0.98 for the proto-Earth's (lower) mantle, corresponding to a predominant abundance of the high-viscosity mineral bridgmanite (Mg,Fe)SiO$_3$, implying poor mixing. While some degree of whole-mantle mixing is indicated by the sinking of a subset of subducted slabs of oceanic lithosphere through the entire mantle, the stagnation of other slab segments\cite{Fukao2013} and the deflection of upwelling plumes\cite{French2015} at $\sim$1000 km depth (i.e. about radius $R$) is indeed consistent with restricted mixing. Sharp seismic-velocity contrasts at similar depths support this interpretation, and provide direct evidence for large-scale compositional mantle heterogeneity\cite{Jenkins2017,Waszek2018}. The preservation of primordial noble gases\cite{Mukhopadhyay2012}, e.g. the large missing budget of argon\cite{allegre1996}, provides complementary evidence for incomplete homogenization of primordial mantle reservoirs. ${}^{182}$W isotopic evidence\cite{Rizo2016,Mundl2017} requires that least a subset of the preserved heterogeneity predates the moon-forming impact, and thus reflects proto-Earth compositions. 

According to our results, the preservation of a significant fraction of primordial mantle heterogeneity through the age of the Earth can explain the isotopic similarity between the Earth's and lunar mantles for a relatively wide range of parent-body compositional differences (Figure \ref{fig:oxygen}). Preservation of SiO$_2$-enriched heterogeneity further helps to balance the bulk-Earth's silica budget relative to the chondritic compositional range. Future geochemical and geophysical studies of lower-mantle composition will contribute to constrain the chemistry and origin of Earth's parent bodies, and thus, ultimately, of the inner solar nebula, providing the means to
test effectively our scenario.



\begin{methods}

  \subsection{Estimation of the penetration depth of the impactor's silicates.}
  In the gravity dominated regime, the interaction between the impactor's mantle and the target's mantle behaves like a fluid collision. One fluid element can only deliver half its momentum/angular momentum to a roughly equal mass fluid element due to the completely inelastic nature of a fluid collision. The transport of angular momentum through shock waves, primarily the contact shock at their first contact, is also inefficient because the shock is almost symmetric to the line of centres. The impactor's mantle keeps roughly half its initial angular momentum, $0.35(1-\gamma)L_{imp}$ (confirmed by our simulations), so that it cannot sink deep into the target's mantle. To avoid rotational instability, the specific angular momentum of the post-impact target must not decrease as radius increases (Rayleigh’s criterion); in simulations, the outer part rotates faster. In the best possible case, we assume the impactor's silicates concentrate in a thin shell and they rotate with materials from the target's mantle, residing outside the shell, at constant specific angular momentum (Extended Data Fig. 1). The materials beyond the shell contain the rest of the angular momentum, $(0.65+0.35\gamma)L_{imp}$. Even for $\gamma=0.15$, the impactor's mantle cannot penetrate half the target's mantle. A hit-and-run collision is more complicated due to interaction with the escaping part and stronger oblique shocks, but the angular momentum transport is still inefficient with $R$ a little deeper (see table 1).
  \subsection{Simulations and analysis.}
  We simulate the giant impacts using the GIZMO code\cite{Hopkins2015}, which is a descendant of the GADGET code\cite{Springel2005} and its SPH method widely used in previous impact simulations). GIZMO indeed contains the legacy SPH implementation of the GADGET code, newer SPH implementations such as PSPH, and, most importantly the novel Meshless Finite Mass (MFM). Gravity is coupled to all these different hydro methods
  using the same treecode scheme \cite{Hopkins2015}.
  The MFM method is an improved hydrodynamics formulation that is fundamentally different from the Smoothed Particle Hydrodynamics (SPH). GIZMO MFM does effective volume partition according to the particle distribution and then solves the Riemann problem to update the fluid variables and can be regarded as a generalized moving mesh method. It employs no explicit artificial viscosity and thus shows better conservation property than SPH\cite{Deng2017a}. MFM captures shocks and subsonic turbulence in giant impact simulations accurately, so it can simulate the mixing properly\cite{Deng2017}. We use about 500K particles in our simulations, which is comparable to present-day high resolution simulations. We run a 2M particles simulation as a convergence test (Extended Data Fig. 2). A comparison between the standard SPH simulation, which suppresses mixing, and MFM simulation is presented in Extended Data Fig. 3.

  We apply the ANEOS/M-ANEOS equation of state\cite{Thompson1974, Melosh2007} with iron comprising the core and dunite comprising the mantle. We build the initial condition for planets (30 wt\% iron, 70 wt\% dunite) by solving the hydrodynamic equilibrium with an isentropic profile and place the particles (computational elements) in spherical shells to represent the equilibrium profile\cite{Reinhardt2017}. The temperature on the planet surface is about 2000K, corresponding to an entropy of 2700 J/kg/K in the mantle and 1200 J/kg/K in the core. We run a comparison study with higher initial entropy and our results are robust concerning the initial entropy value (Extended Data Fig. 4). Note that the same version of ANEOS/M-ANEOS is implemented in
  SPH and MFM, so that any difference that we observe between simulations carried out with these two methods will only stem from the underlying hydrodynamical solver.

  We characterize the modeled impacts by determining the disk mass/angular momentum, the predicted Moon mass, planet mass/angular momentum and the internal structure of the post-impact target. The former is done following a standard approach, i.e., bounded particles with periapsis distance larger than the equatorial radius of the planet are classified as disk particles\cite{Canup2013,Rufu2017}. This calculation is preformed at least 40 hours after the impact, i.e. when the system saturates to a quasi-steady state \cite{Canup2013} (Extended Data Fig. 5).  We calculate the entropy profile by arithmetically averaging the entropy across spherical shells (assuming spherical symmetry). The fraction of materials from the target is calculated analogously. The iron core is slightly oblique with an equatorial radius that is slightly ($< 5\%$) larger than the polar radius; hence, the profiles is not messed up at the core-mantle boundary. A counter-intuitive result involves that that core entropy can drop below its initial value for MFM, as shown by the entropy profile (e.g., Figure 1d).
  
 This prediction by our MFM models contrasts with that of our SPH models (Extended Data Fig. 3). The entropy drop in the core for MFM is caused by a phase transition and the associated redistribution of internal energy, and entropy, from the core to the mantle, hence it has a physical origin\cite{Deng2017}. Indeed, as the outer core, near the core-mantle boundary, melts first, entropy locally increases, forcing the inner core to decrease its own entropy in order to maintain thermodynamical equilibrium. This entropy loss outweighs the entropy gain through shocks in the central core region. In contrast, in SPH simulations, the core and mantle are separated by an artificial tensional force \cite{Agertz2007,Hosono2016,Deng2017} (see also Extended Data Fig. 6). This force largely isolates the core, causing the core to evolve nearly adiabatically\cite{Deng2017}. The physical heating associated with shocks is somewhat overestimated by the usage of artificial viscosity\cite{Springel2005,Deng2017}; as a result, the entropy slightly increases in the core for SPH. That the entropy in the central core does not decrease for SPH, is thus mostly due to a numerical artifact. The marked difference between MFM and SPH, i.e. in terms of the temperature and entropy distribution near the core-mantle boundary, is visualized in Extended Data Fig. 6.

  \subsection{Code availability}
  The latest version of the GIZMO code is made available by its author, Philip Hopkins at http://www.tapir.caltech.edu/~phopkins/Site/GIZMO.html
  \subsection{Data availability}
  The data files that support our analysis will be made available upon reasonable request.
  
\end{methods}




\begin{addendum}
 \item We thank Martin Jutzi, Prasenjit Saha, Romain Teyssier and Qing-Zhu Yin for stimulating discussions. We acknowledge support from the Swiss National Science Foundation via the NCCR PlanetS.
 \item[Contributions] H.D. conceived the idea of linking the moon formation impact to the Earth mantle heterogeneity and planned the project. C.R. built the equation of state (EOS) library and F.B. prepared the EOS lookup table. H.D. and C.R. incorporated the EOS library into the hydrodynamical code and prepared the initial conditions. H.D. ran the simulations and did the visualisation and interpretation. C.R., L.M., J.S. also helped in the interpretation. M.B. and M.M. contributed to the geodynamic and cosmochemistry argument, respectively. H.D., M.B. and M.M. prepared the manuscript. L.M. reviewed the manuscript and all authors commented on it.
 \item[Competing Interests] The authors declare that they have no
competing financial interests.
 \item[Correspondence] Correspondence and requests for materials
should be addressed to H.D.(email:hpdeng@physik.uzh.ch)
\end{addendum}


\bibliographystyle{naturemag}
\bibliography{references}

\begin{table}
\footnotesize
\caption{Parameters and results of impact simulations\label{tab:simulations}}
\medskip
\begin{tabular}{@{}llllllllllllll}
\hline
Run &  $M_{tar}$ & $M_{imp}$ & $b$ & $\frac{v_{imp}}{v_{esc}}$ & $\frac{L_{F}}{L_{EM}}$ & $\frac{L_D}{L_{EM}}$ & $\frac{M_{D}}{M_L}$ & $\frac{M_{planet}}{M_{E}}$ & $\frac{M_U}{M_{mantle}}$ & $F_{U,tar}$ & $F_{D,tar}$ & $\frac{M_{Fe}}{M_D}$ & $\frac{M_M}{M_L}$\\
\hline
1   &0.85     & 0.12 &  0.71       & 1.00               & 0.91             & 0.16            &  0.87              &  0.94                       &0.46   &0.84  &0.38   &0.16   &0.73  \\
\textcolor{red}{2}   &0.85     & 0.12 &  0.73       & 1.00               & 1.18             & 0.35            &  1.48              &  0.95        &0.50     &0.86  &0.40   &0.04   &1.48\\
\textcolor{red}{3}    &0.85     & 0.16 &  0.71       & 1.00               & 1.30             & 0.32            &  1.70              &  0.98         &0.49   &0.80  &0.30   &0.04   &1.5 \\
  4   &0.85     & 0.16 &  0.73       & 1.00               & 1.30             & 0.23            &  1.22              &  0.98                      & 0.43  &0.79  & 0.39  &0.12   &1.08 \\
\textcolor{red}{5}   &0.90     & 0.16 &  0.71       &1.00                & 1.32             & 0.32            &  1.98              &  1.0         & 0.52  &0.81  &0.47   &0.01   &1.19   \\
\textcolor{red}{6}    &0.90     & 0.16 &  0.73       &1.00                & 1.53             & 0.38            &  1.64              &  1.0          & 0.52  &0.80  &0.34   &0.05   &1.64 \\
\hdashline
\textcolor{red}{7}   &0.85     & 0.2  & 0.574       &1.20               &1.33              &0.29             & 1.63                & 0.99            & 0.59  &0.79  &0.52    &0.08   &1.26  \\
\textcolor{red}{8}   &0.85     & 0.2  & 0.574       &1.25               &1.26              &0.26             & 1.45                & 0.98             & 0.62  &0.80  & 0.47   &0.09   &1.14   \\
  9   &0.90     & 0.2  & 0.537       &1.20               &1.30              &0.05             & 0.30                & 1.06                          & 0.55  &0.79  & 0.64   &0.01   &0.20 \\
  10  &0.90     & 0.2  & 0.537       &1.25               &1.32              &0.13             & 0.81                & 1.04                         & 0.60  &0.79  & 0.62   &0.03   &0.48   \\
  11  &0.90     & 0.2  & 0.537       &1.30               &1.30              &0.20             & 1.20                & 1.04                         & 0.54  &0.79  & 0.55   &0.12   &0.79  \\
  12  &0.90     & 0.2  & 0.574       & 1.15              &1.38              &0.14             &0.88                 & 1.04                        & 0.54  &0.79  &0.59    &0.01   &0.51  \\
\textcolor{red}{13}  &0.90     & 0.2  & 0.574       & 1.20              &1.34              &0.23             & 1.23                & 1.02           & 0.56  &0.79  &0.57    &0.08   &1.07  \\
\textcolor{red}{14}  &0.90     & 0.2  & 0.574       & 1.25              &1.30              &0.24             &1.30                 & 1.02           & 0.59  &0.80  &0.54    &0.09   &1.30  \\
  15  &0.90     & 0.2  & 0.574       & 1.30              &1.20              &0.23             &1.18                 & 1.02                  & 0.50  &0.79  &0.44    &0.16   &1.18  \\
\hline
\end{tabular}
At the end of each run (at least 40 hours after the impact, until no clumps present in the disk) we evaluate the proto-lunar disk and the post-impact target compositions following an established approach\cite{Rufu2017,Canup2013} (see Methods subsection 2). The canonical (top) and hit-and-run (bottom) models are separated by the horizontal dashed line. $M_E, M_{L},M_{D}, M_{M}, M_{planet}$ are the real masses of the Earth and Moon, and the predicted masses of the proto-lunar disk, the formed moon, and post-impact target (formed Earth), respectively. We also calculated the mass fraction of the mantle that lies beyond $R$, i.e., $\frac{M_U}{M_{mantle}}$. The fraction of target's silicates in the upper layer mantle ($r>R$) is denoted as $F_{U,tar}$. Runs with $M_M>1.0M_L$ and $M_{Fe}/M_{D}<0.1$ are regarded as successful impacts and highlighted with the red color\cite{Cuk2012}.
\end{table}

\begin{figure}
  \centering
  \includegraphics[scale=0.6]{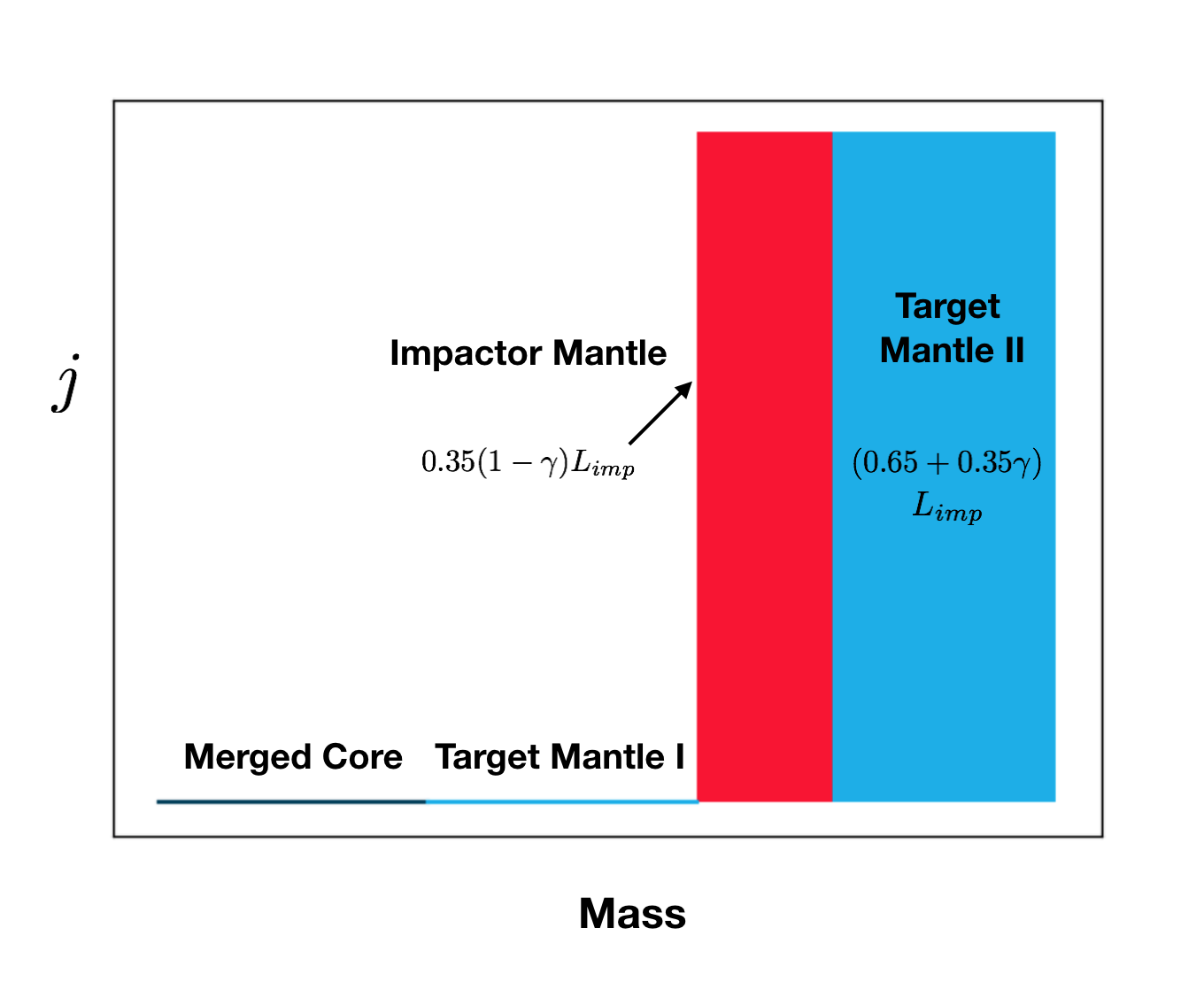}
  \caption{\textbf{Extended data figure 1}: An idealized model to estimate the maximum possible penetration depth. The specific angular momentum profile, which must not decrease with radius to avoid rotational instability, is just a step function. The silicate materials can be placed outside the impactor's silicates are less than those inside the impactor's silicates, thus smaller than half the target mantle's mass for an impact with $\gamma<0.15$.\label{extfig:1}}
\end{figure}

\begin{figure}
  \includegraphics[width=\textwidth]{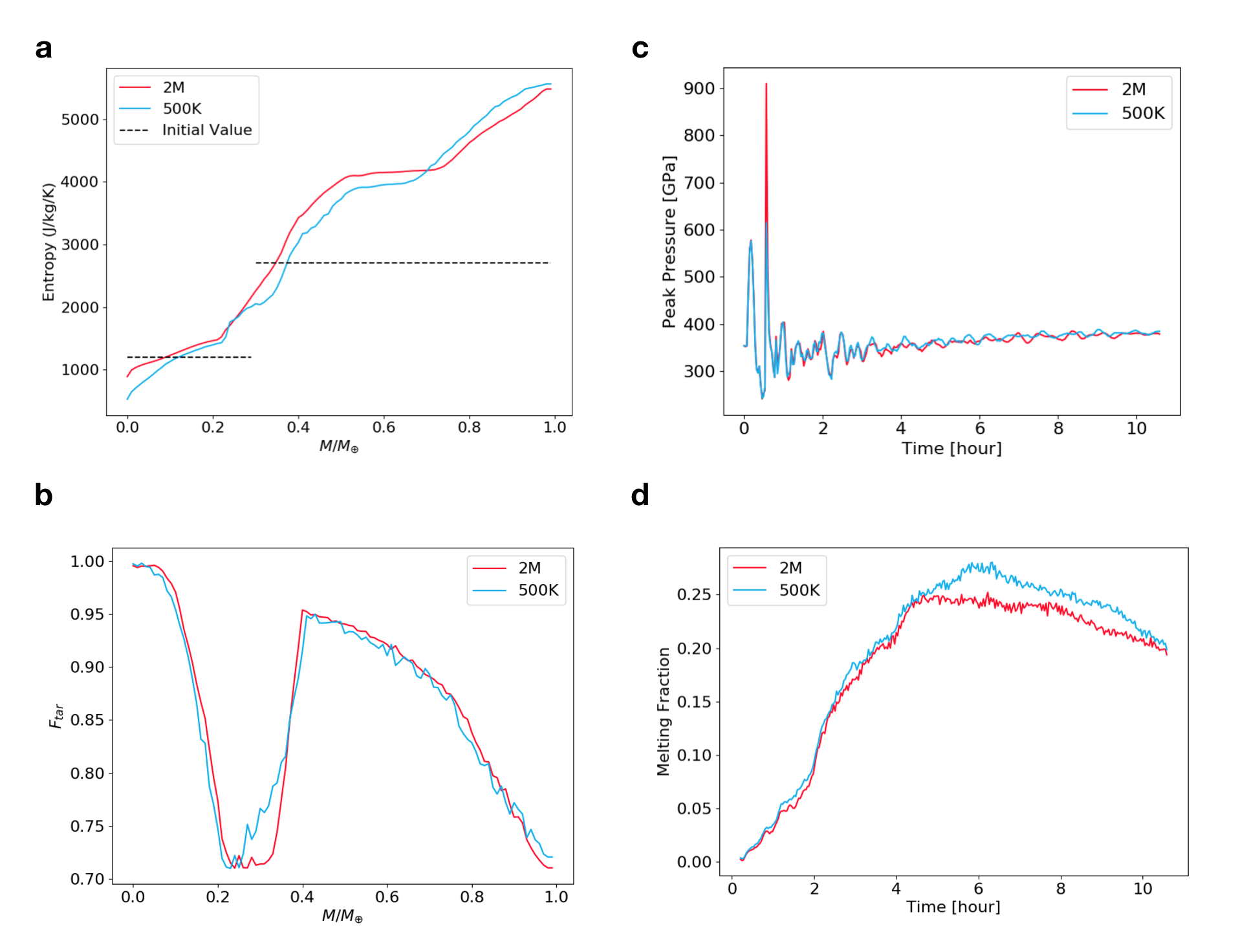}
   \caption{\textbf{Extended data figure 2}: Resolution test of our MFM simulations in terms of number of particles. A high-resolution case with 2M particles (and otherwise the same parameters as run 13) indeed numerically converges well with run 13. Even though small differences persist in terms of the details of the (\textbf{a}) entropy profile, (\textbf{c}) peak pressure and (\textbf{d}) melt fraction of the core\cite{Pierazzo1997}, the most critical model prediction (i.e., (\textbf{b}) the compositional profile) converges well at the two resolutions shown. The pressure fluctuation is stronger in the lower resolution simulation (due to larger discretization noise), which leads to exaggerated melt fraction in the core.\label{extfig:2}}
\end{figure}
\begin{figure}
  \centering
  \includegraphics[scale=0.4]{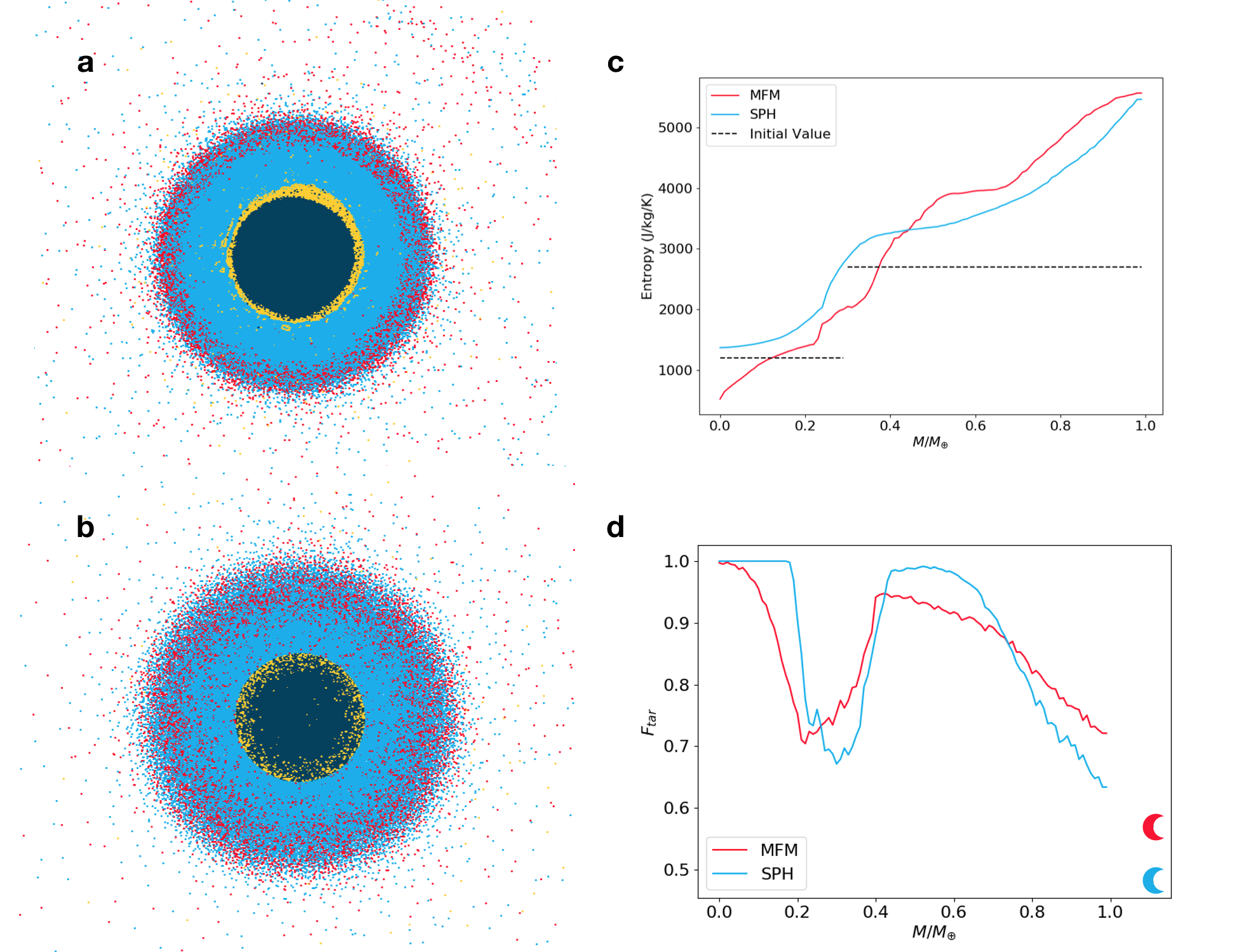}
  \caption{\textbf{Extended data figure 3}: Comparison between MFM and standard SPH simulation of run 13 in table 1. \textbf{a} and \textbf{b} show the material distribution in the post-impact target in the SPH and MFM run respectively. Compared to MFM, impactor materials tend to be enhanced near the planet's surface and near to top of the core in SPH, and nearly absent in the deep mantle and core\cite{Deng2017}. \textbf{c} The entropy profile for both methods. The entropy jump in the mantle is sharper for MFM than for SPH. In the MFM simulations, the entropy drop in the core is explained by transfer of energy due to the phase transition at the core-mantle boundary\cite{Deng2017} (Extended data figure 6). \textbf{d} The mass fraction of materials from the target is plotted as a function of the normalized enclosed mass. Mixing of impactor material with the deep target mantle and core is more efficient for MFM than for SPH. For a detailed discussion in terms of the comparison of both approaches, we refer the reader to \cite{Deng2017}.\label{extfig:3}}
\end{figure}
\begin{figure}
  \centering
  \includegraphics[scale=0.6]{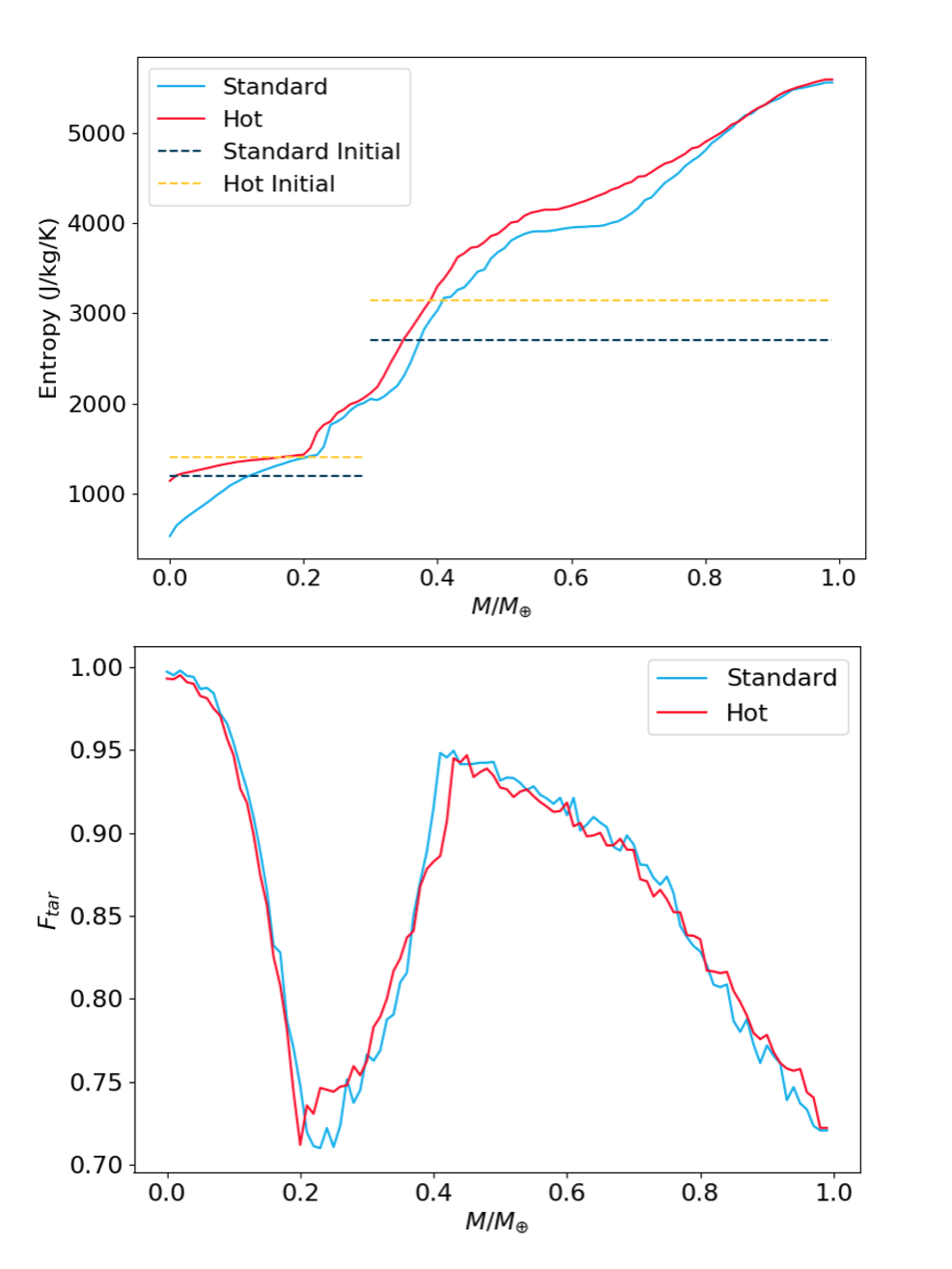}
  \caption{\textbf{Extended data figure 4} Model results as a function of initial condition. The upper and lower panels shows the final entropy and compositional profiles, respectively, for run 13 (blue), and an analogous case with a higher initial entropy target (1400 J/kg/K for iron and 3200 J/kg/K for dunite) (red). In the case with an initially hotter target, the entropy jump in the mantle is slightly less sharp than for run 13. However, the compositional profile is robust, remaining virtually unchanged. The disk mass and angular momentum also remain robust with differences $<$2\%.\label{extfig:4}}
\end{figure}

\begin{figure}
  \centering
  \includegraphics[scale=0.6]{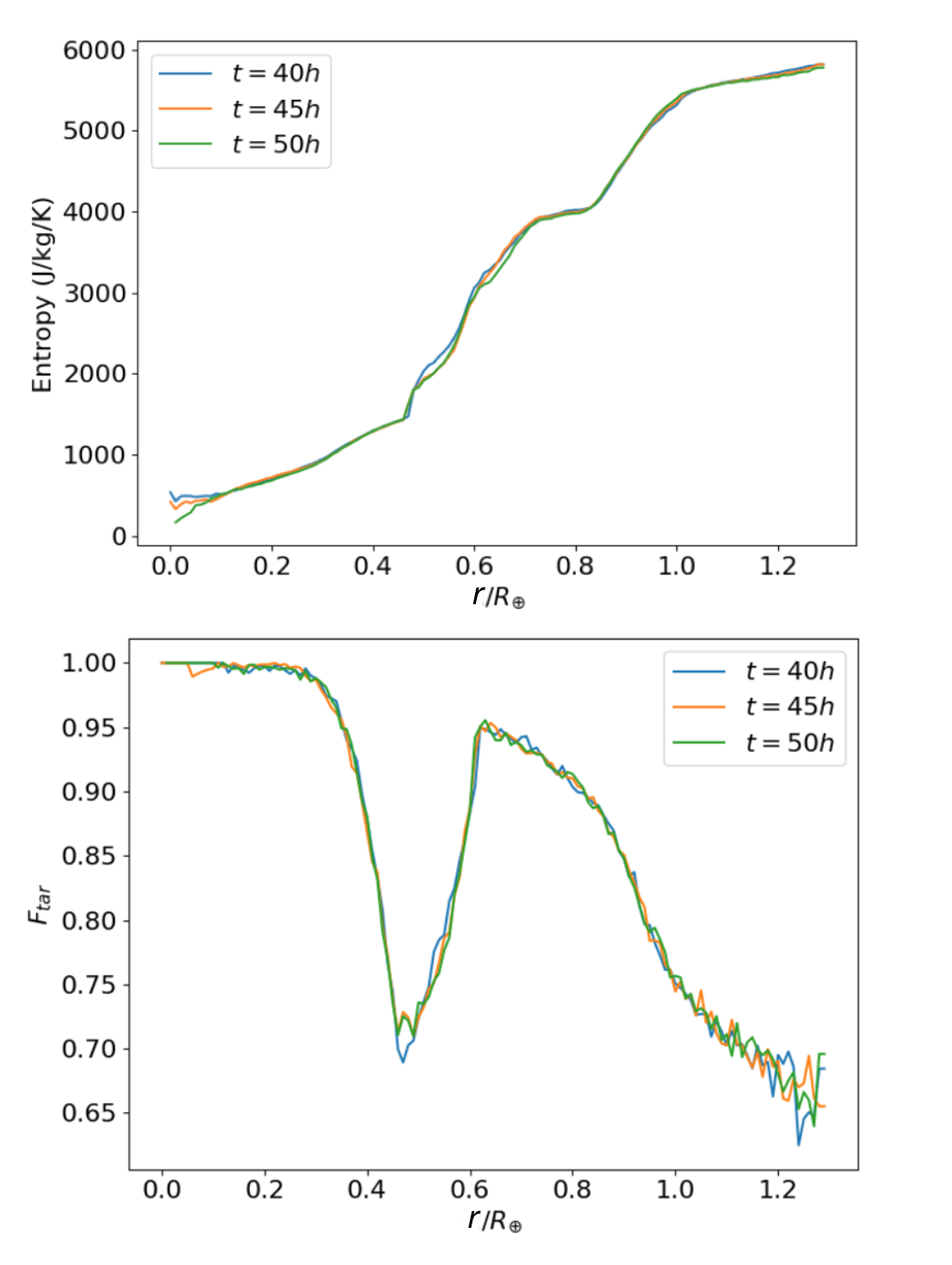}
  \caption{\textbf{Extended data figure 5} The saturation of the thermal and compositional state of the post-impact target. The upper and lower panels display the entropy and compositional profiles, respectively, at different model times for run 13. After 40 hours, the post-impact target already has reached a stble state. This result confirms that our approach of performing the analysis at $\geq$40 hours after the impact is reliable. Here is the movie for \href{https://drive.google.com/open?id=1pZQC3aX_wgDiwLhTi-XUKG3Z07lbsISR}{run 13} (note the code time unit is 1.77 h).\label{extfig:5}}
\end{figure}

\begin{figure}
  \includegraphics[width=\textwidth]{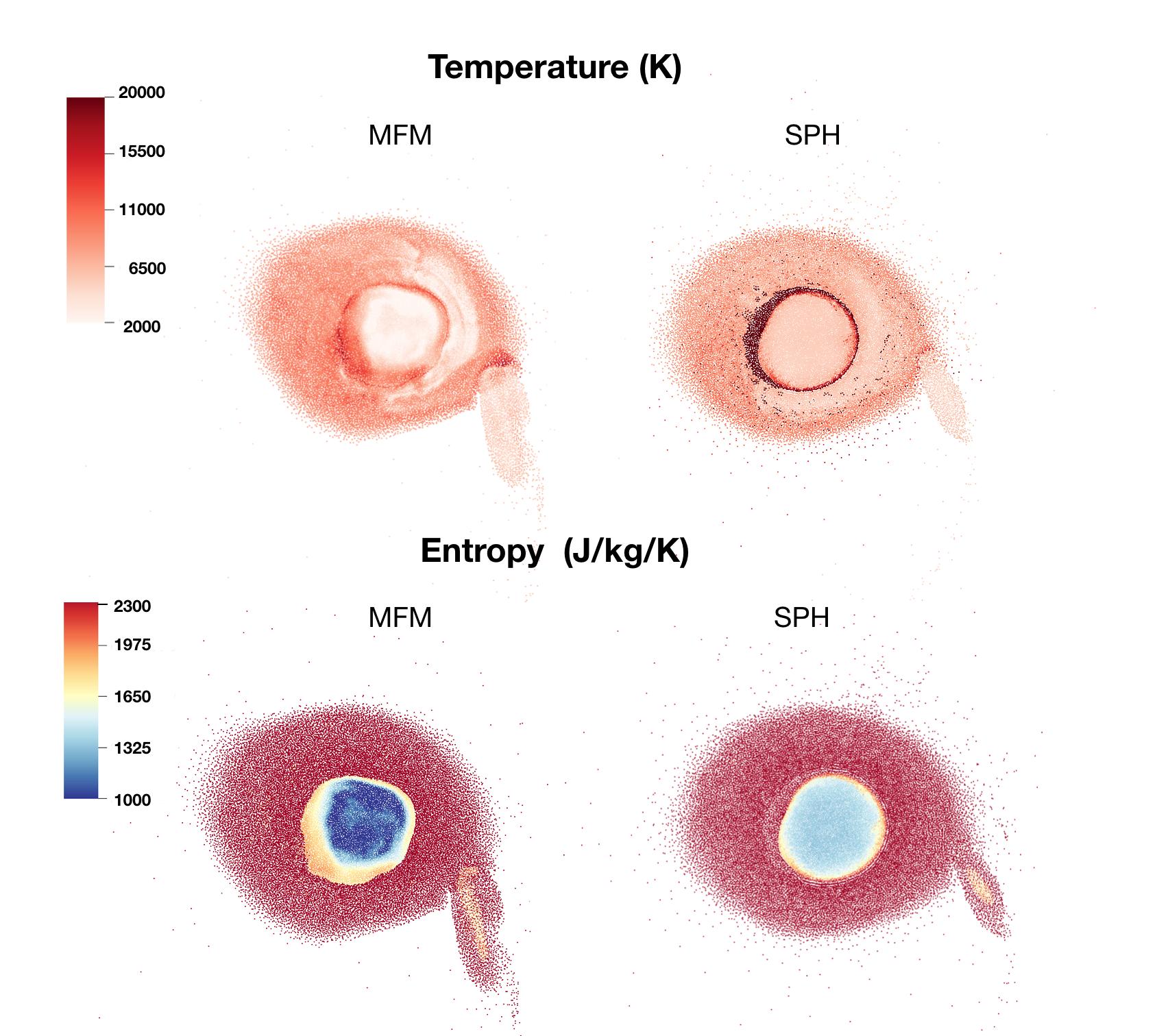}
  \caption{\textbf{Extended data figure 6} The difference of the thermal state in MFM and SPH. We show snapshots of the SPH and MFM comparison run at 7.08 hours. There is a clear separation between the core and mantle in the SPH simulation due to numeric issues\cite{Deng2017}; the temperature difference across the core-mantle boundary can be larger than 10000K.\label{extfig:7}}
\end{figure}



\includepdf[pages=-]{empty.pdf}
\includepdf[pages=-]{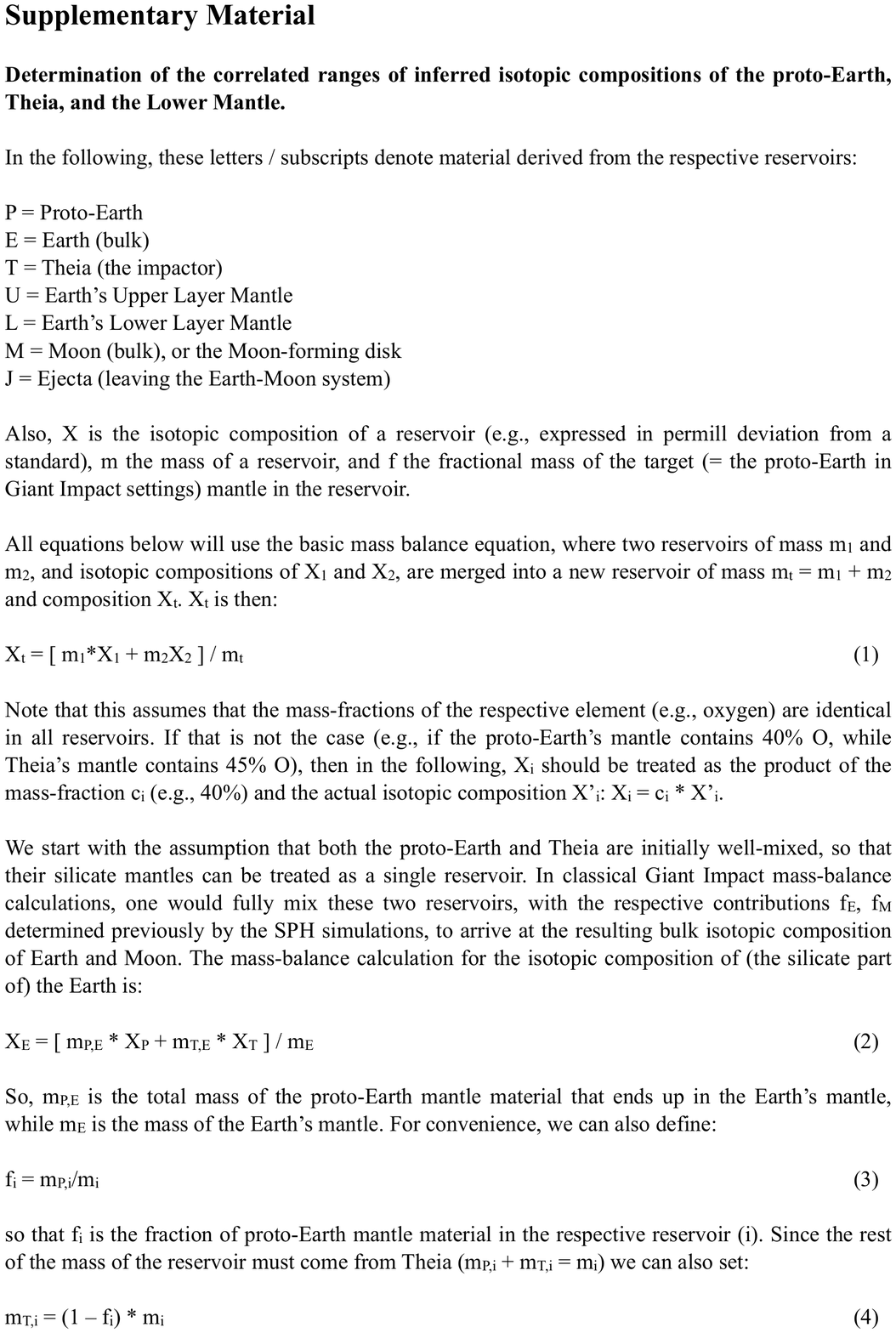}
\includepdf[pages=-]{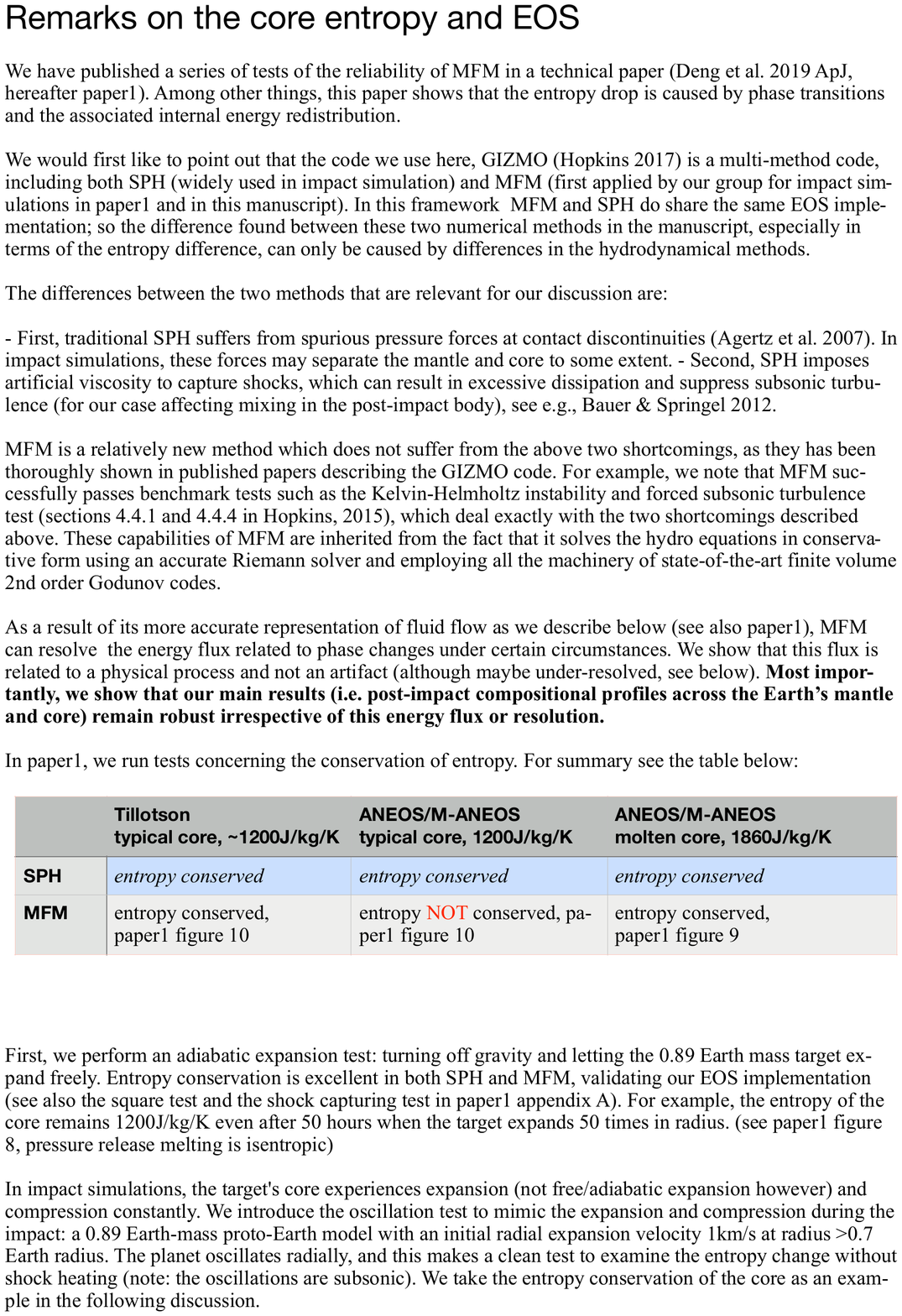}
\end{document}